# GOVERNMENT LICENSE

# Operating advanced scientific instruments with AI agents that learn on the job

Aikaterini Vriza[1,*], Michael H. Prince[2], Tao Zhou[1], Henry Chan[1], Mathew J. Cherukara[2,*]

[1] Center for Nanoscale Materials, Argonne National Laboratory, IL, USA
[2] Advanced Photon Source, Argonne National Laboratory, IL, USA

## Abstract

Advanced scientific user facilities, such as next generation X-ray light sources and self-driving laboratories, are revolutionizing scientific discovery by automating routine tasks and enabling rapid experimentation and characterizations. However, these facilities must continuously evolve to support new experimental workflows, adapt to diverse user projects, and meet growing demands for more intricate instruments and experiments. This continuous development introduces significant operational complexity, necessitating a focus on usability, reproducibility, and intuitive human-instrument interaction. In this work, we explore the integration of agentic AI, powered by Large Language Models (LLMs), as a transformative tool to achieve this goal. We present our approach to developing a human-in-the-loop pipeline for operating advanced instruments including an X-ray nanoprobe beamline and an autonomous robotic station dedicated to the design and characterization of materials. Specifically, we evaluate the potential of various LLMs as trainable scientific assistants for orchestrating complex, multi-task workflows, which also include multimodal data, optimizing their performance through optional human input and iterative learning. We demonstrate the ability of AI agents to bridge the gap between advanced automation and user-friendly operation, paving the way for more adaptable and intelligent scientific facilities.

**Keywords:** Multiagent workflows, Self-driving Laboratories, Large Language Models, X-ray scattering, Synchrotron, User facilities

[*] kvriza@gmail.com, mcherukara@anl.gov

# 1. Introduction

Scientific user facilities, such as synchrotron light sources, ultrafast lasers and increasingly self-driving laboratories are transforming experimental science by automating routine tasks and decreasing the time required to go from hypothesis to result.[1–3] The emergence of Foundation Models (FMs), i.e., large language, vision and multimodal networks pre-trained on vast amounts of data, is reshaping how scientists can interact with advanced user facilities, enabling safer, faster and more efficient laboratory operations.[4–6] While significant progress has been made in adapting FMs for various scientific applications,[7,8] deploying them as autonomous agents for real-world laboratory experimentation remains elusive.[9] The design, synthesis, characterization, and testing of new materials often entails a complex, multi-step sequence of actions that span from instrument control to data interpretation. This requires not just robust coding frameworks but also a deep, often visual, understanding of the experimental outputs. Although automation has made significant progress in executing individual protocols and automating experimental procedures, achieving true autonomy in scientific discovery involving tasks such as experimental workflow design, data analysis, execution planning, and on-the-fly, decision making remains an active challenge that will likely continue to require human feedback and expert knowledge.[10,11]

Tasks that human scientists perform intuitively, such as identifying bright spots in a 2D scan or manipulating laboratory equipment, present significant challenges to autonomous systems. For example, picking up and moving a vial with a robotic instrument requires precise coordination of multiple components like grippers and vial holders, along with accurate spatial awareness. Traditional robotic systems rely on predefined functions for these operations, limiting their flexibility and ability to adapt to different experimental environments. Moreover, while existing LLM agents can utilize predefined functions as tools, they often lack the capability to incorporate user feedback and autonomously design experimental processes. Significant progress has been made in the deployment of LLMs in scientific research, from decision-making and literature screening to the execution of experimental protocols.[12–14] Research teams have successfully expanded LLM capabilities by integrating them with diverse sets of tools.[15,16]

The field is now advancing towards collaborative multi-agent systems that enhance the capabilities of LLMs through iterative feedback that can be provided via audio, text or video/imaging, role specialization, and coordinated teamwork.[16–19] These systems draw inspiration from human scientific discovery processes and are designed to complement human creativity and expertise with AI's ability to analyze vast datasets, navigate hypothesis spaces and execute repetitive tasks. The adoption of pre-built frameworks such as LangGraph[20], Autogen[21] and Swarm[22] has significantly streamlined the integration of LLMs with scientific workflows, enabling efficient coordination between LLM agents and external tools.[23] LLM-powered agents that combine text generation with decision making, memory and tool execution to autonomously perform tasks in iterative workflows and can impact areas across various scientific disciplines.[24] AI agents can impact areas across various scientific disciplines. In biology, for instance, biomedical AI agents are being applied to areas such as virtual cell simulation, programmable control of phenotypes, cellular circuit design, and the development of novel therapies.[25] In materials science, platforms like AtomAgents, are being developed as agentic systems for knowledge retrieval, multi-modal data integration, physics-based simulations, and comprehensive results analysis.[26]

In this work, we develop and benchmark a multi-agent framework that operates complex scientific instrumentation at two distinct scientific user facilities, a synchrotron X-ray nanoprobe beamline and an autonomous robotic station for iterative materials design. The agents can orchestrate multistep experimental workflows, interpret multimodal data streams and interactively collaborate with human researchers. An important aspect of this study is the ability of the system to be learn on the job by integrating expert guidance through in-context learning. We integrate this learning in two ways. For the highly standardized X-ray experiments, the vision agent is provided with expert-designed prompt templates that encode scanning and diffraction analysis heuristics. For the more open-ended robotic workflows, users can explain the operational procedures in real time and each instruction is automatically embedded as a user-interaction memory that the agent can retrieve in future sessions. We evaluated the agents' ability to learn from human scientists and their potential for self-improvement through these interactions.

## 2. Results

### 2.1. Overview of the agentic pipeline

The core of our agentic pipeline is the Autogen (AG2) framework[21] with integrated LLMs with vision capabilities. Our proposed workflow is separated in three main levels, i) the human level, where human users can provide instructions or related literature for reference, ii) the AI agent level, which is an orchestration of several agents with access to data, memories from previous interactions and operational protocols of the laboratory equipment, iii) The physical instruments level, where the agents interact with the existing equipment through an executor which can operate the instruments and provide feedback to the agents for the successful operation or any errors during the execution (Fig. 1). The specialized agents are presented in detail in Table 1. The AI agents have access to the Hard X-ray Nanoprobe (HXN) and N9 robot's operation commands, station layout, and task descriptions, which are provided as a context. The pipeline also leverages built-in capabilities, such as linter tool for code checking and the executor platform for running the generated code through the administrator agent.

A central feature of our workflow is the ability of the agents to learn on the job, i.e. continuously learn from human feedback. In this approach, agent capabilities are updated by appending a new instruction to the system message, integrating storage and retrieval functionalities, and interacting with a background agent that decides whether the human provided information is important and should be stored. In that way, any new interaction with the human researcher that provides significant guidelines for the execution of the experimental protocol is stored as an input-output pair in a local vector database. This process enables the agents to recall and apply previously learnt information when facing similar tasks, thereby supporting long-term learning and adaptability. See Methods 4.2 for additional details and Supplementary Section 3, Fig. 8 and 9.

**Table 1**: Available agents and their operating capabilities. The agents have access to operation commands of the robotic systems and station layout.

| Agent | Agentic Tasks |
|---|---|
| Code writer | Writes the code after being provided with the relevant system files. |
| Code critic | Checks the code written from the code writer and provides feedback. |
| Administrator | Interacts with the human user and the other agents. Executes the code in the robotic working environment and interacts with the human for feedback. |
| Paper scraper | Scrapes research articles for relevant information extraction. |
| Image explainer | Image analysis agent. When provided with an image, identifies relevant details and provides comprehensive answers. |
| Teachability | Memory retrieval of previously saved interactions between humans and agents. |

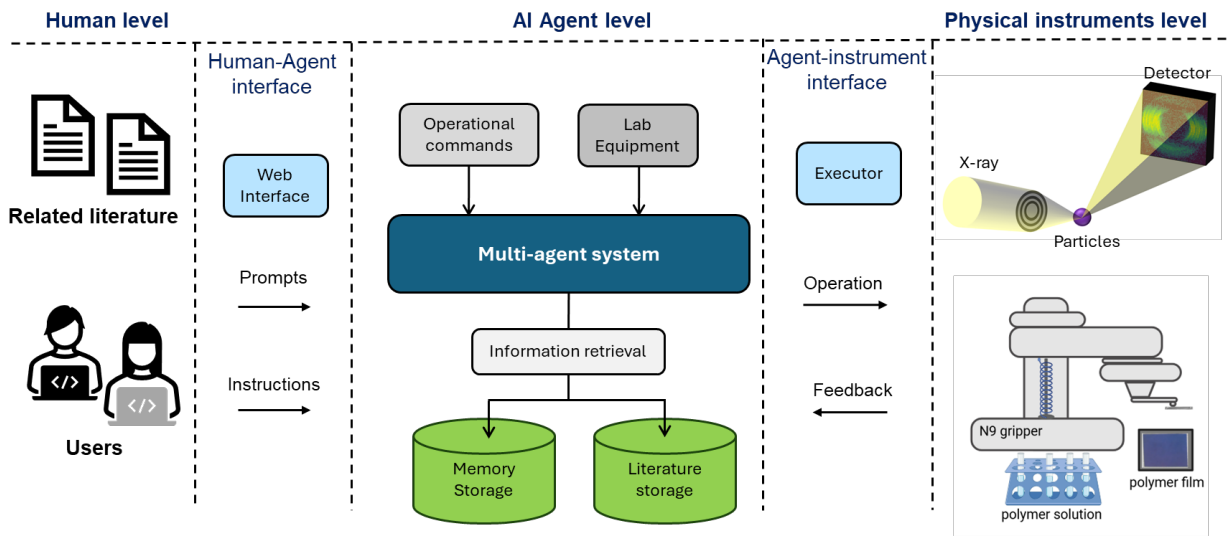

**Fig 1. Outline of the proposed agentic workflow**. The system is initiated by facility users, who can optionally provide related reference papers to the agents and define tasks via prompts. These prompts may include details about the main equipment components and relevant system files, which are provided as context to the agentic pipeline. The pipeline consists of built-in specialized agents; each integrated with external tools and memory. The physical instruments demonstrated in this work include the X-ray Nanoprobe Instrument for nanoscale imaging and a materials design robot for thin film design and processing. Interaction between the agents and instruments is facilitated via an executor which submits workflow scripts to the instrument's working environment.

## 2.3. X-ray Nanoprobe Experiment

### 2.3.1 Hard X-ray Nanoprobe Instrument

The HXN beamline uses motorized and robotic control to perform strain imaging of functional materials at the nanoscale.[27] Because of radiation hazards, no personnel are allowed in the experiment hutch while the beam is active. As such, the entire instrument is controlled remotely via the beamline customized software. The backbone of the control interface is the Experimental Physics and Industrial Control System (EPICS) software and is operated through predefined Python functions and the PyEpics library (Supplementary Fig. 1). An outline of the existing beamline system is presented in (Fig. 2) and includes sample and optics motors as well as X-ray fluorescence and diffraction detectors. The main operational commands involve defining the motors and the executable helper functions as Python commands in a single python script (Supplementary Fig. 2). A typical operation begins by identifying the motors and layout that should be used to control the beamline. Then the pre-defined functions are used to either move the beam or the sample to a user selected location and conduct a two-dimensional scan.

The typical two-dimensional map command requires the user to input four sets of information, the name of the two motors, their start and end motor positions, the numbers of steps and the exposure time. The first motor to appear in the command is the outer loop motor while the second motor is the inner loop motor. The inner loop motor scans from start to end, then the outer loop motor steps forward by one position. This process repeats—scanning the inner loop for each step of the outer loop—until the outer loop completes its full range, generating a 2D map. The first step is typically to perform a survey scan with coarse motor steps (spatial resolution) and short exposure time. The users then look at the integrated intensity of the survey scan to pick an area of interest to perform more detailed investigations. To replicate these tasks through the agentic pipeline we employed the code-writer agent, the code-reviewer agent and the image-explainer agent. The main tasks that were assigned to the agents range from function execution using either logical or algebraic inference to acquire a 2D scan to the more advanced image analysis to identify the correct coordinates to zoom and get a new scan.

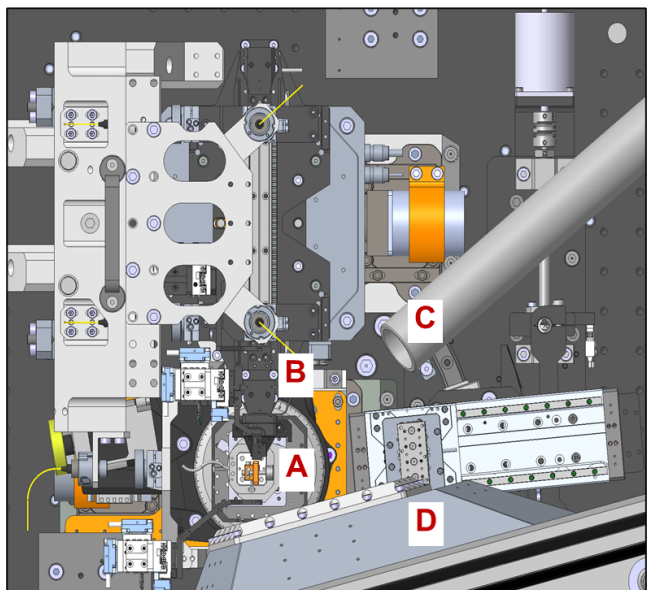

**Equipment**

A. Sample scanning motor stack
B. Optics scanning motor stack
C. Towards X-ray fluorescence detector
D. Towards X-ray diffraction detector

**Examples of operational commands**

```
# Define motors
zpx = epics.Device('26idcnpi:X_HYBRID_SP.',
attrs=('VAL','DESC')) # motor for moving the x-ray beam in the x
direction. A positive value moves the beam outboard.

# Define operational functions
def fly2d(motor1, startpos1, endpos1, numpts1, motor2, startpos2,
endpos2, numpts2, dettime, absolute=False):
# scan two motors for a mesh scan
# the unit of the scan is in um
# motor1 is the outer loop motor# motor2 is the inner loop motor

def mov(motor,position):
    # move motor to absolute position
```

|       | Tasks | Complexity |
|-------|-------|------------|
| Task1 | Take a scan in a 100 um x 100 um area, with 1 um resolution, 0.01 sec exposure time. | Easy |
| Task2 | In a single particle image of the scan, identify the location of the particle and move the beamline there. | Medium |
| Task3 | Compare a multi-particle nano-diffraction image with and the corresponding nano-fluorescence image to identify the best location to move the beam by selecting a bright and well isolated point. | Hard |

**Fig 2. X-ray nanoprobe robotic system**. The system includes sample and optics scanning motors and X-ray fluorescence and diffraction detectors. The main operational instructions involve defining the motors and executable functions as Python commands. The table describes the tasks of increased complexity that were used to evaluate the performance of multimodal large language models within the agentic pipeline for X-ray imaging operations.

### 2.3.2 Agentic pipeline evaluation at the Hard X-ray Nanoprobe

For the Hard X-ray Nanoprobe (HXN) beamline at the Advanced Photon Source (APS) (Fig. 2), the primary focus is coordinating a vision agent that can identify the positions of suitable nanoparticle candidates in the images with the code writer and code reviewer agents that will translate the visual observations into actionable tasks and operate the instrument.

Fig. 3a presents execution results for two representative tasks of increasing complexity (detailed results can be found in Supplementary Section 1.1, Supplementary Fig. 4 and 5). In the first example the user provides a minimal prompt requesting a coarse survey scan over a specified range with 1 um resolution and 0.01 sec exposure time. However, no start and end positions are given, and the agent should infer them based on the scan range provided. The code writer agent should generate the correct command and translate the user prompt into the correct input parameters for the function.

The second example illustrates a more complex scenario involving the integration of a vision agent into the agentic environment. Here, the vision agent is responsible for interpreting and analyzing experimental images to assist with decision making. A key challenge addressed in this task is the identification of optimal scanning regions through a cross-modality analysis of nano-diffraction and nano-fluorescence images (Fig. 3b). In the nano-diffraction images, desirable regions contain isolated bright spots, which correspond to particles that diffract under the given diffraction conditions. These are the particles of interest for subsequent measurements. The brighter the spot, the better for further analysis. In contrast, nano-fluorescence images tend to display broader clusters of bright areas, caused by the coalescence of multiple particles. These bright areas that look like clusters of bright spots of more than 10 um x10 um should be avoided to ensure subsequent measurements only on single, non-clustered particles. The agents were provided with instructions on how to get visual insights from the nano-diffraction and nano-fluorescence images as context (Supplementary Fig. 3).

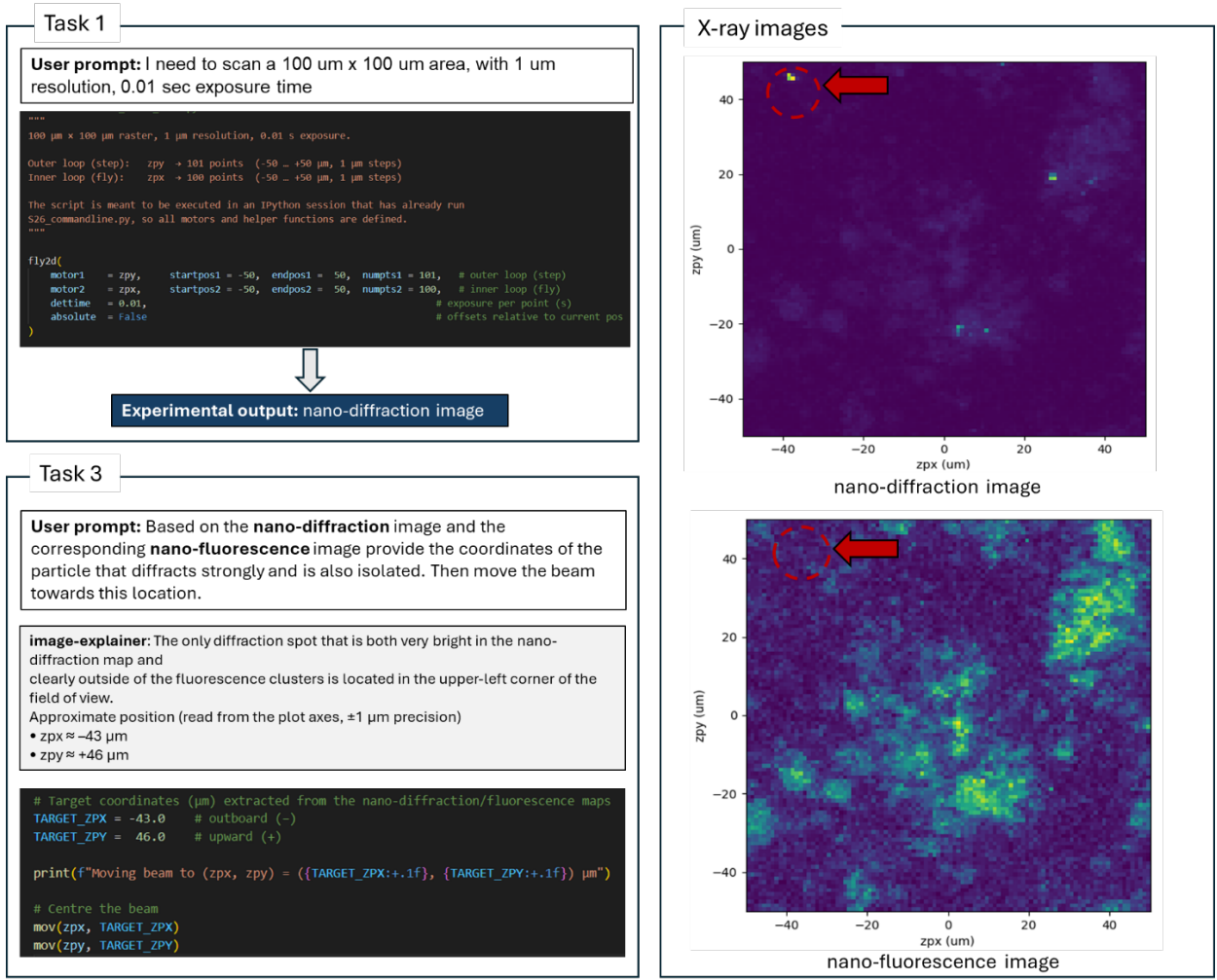

**Fig 3. Operational examples of the X-ray nanoprobe system using the agentic pipeline**. The operation is initiated with the user input aiming to perform specific scans. a) Two examples demonstrating the workflow. As a first example the user wants to perform a scan in a specific area with certain resolution and exposure time. The code-writer agent using the advanced reasoning o3 model identified the correct function with the desired parameters and the executor performed the action for getting the scan. In the second most complicated example, after the first scan is taken, the user needs to identify the optimal location for scanning via a cross-modality check. The optimal location is defined by a region with bright spots in the nano-diffraction image which corresponds to a region which is more isolated in the nano-fluorescence image. By performing a cross-modality check the vision agent based on o3 model was able to correctly identify the optimal region with high precision on the coordinates. The cross-modality check involves. b) The nano-diffraction (top) and the nano-fluorescence (bottom) images used for the cross-modality check.

In this case, the image-explainer agent was provided with a nano-diffraction image that had visually three distinct bright spots. The coordinates of three areas should be identified correctly by interpolating the axes coordinates. When the brightest positions are found on the diffraction image which precision of 1 um, a cross-modality check should be performed to the corresponding nano-fluorescence image to select the coordinates that correspond to the least clustered region. Among the models tested (Fig. 4), only the o3 model consistently demonstrated the ability to reason across modalities with high positional precision. It correctly extrapolated from image axes and maintained consistent behavior across trials. While gpt-4o exhibited strong performance in language-based instruction tasks, especially when fine-tuned with human feedback, it was less reliable in multimodal reasoning. Both gpt-4o and gpt-4o-mini struggled to interpret axis values and cross-reference spatial information across image types, resulting in lower performance for image-grounded tasks. Moreover, o3 showed the highest consistency in the output among all tasks, with Claude 3.5 having the most inconsistent results through the trials (Supplementary Fig. 10).

Notably, the o3 model also provided more coherent reasoning for parameter selection in both text-only and image-guided tasks. In the first task, although o3 occasionally hallucinated code, its integration with a code-reviewer agent helped ensure correctness, ultimately improving the reliability of the code-writing agent. Overall, o3 demonstrated superior capability in handling both textual and visual modalities, making it particularly effective for agent-based scientific workflows involving complex, multimodal reasoning. A key observation on the teachability aspect is that while human feedback improves performance in text-based tasks such as function calling or it has limited impact on enhancing the models' inherent visual reasoning capabilities, e.g. the feedback fails to compensate for deficiencies in interpreting and extrapolating from image-based data. So, for the two tasks (2 and 3) that involve identifying points in the scan and calling the function with correct parameters, providing human feedback only improves the agent's understanding of the correct function calling and not their understanding of the image. The comprehensive evaluation of the tasks is provided in the Supplementary Section 4.1, Supplementary Tables 1-6.

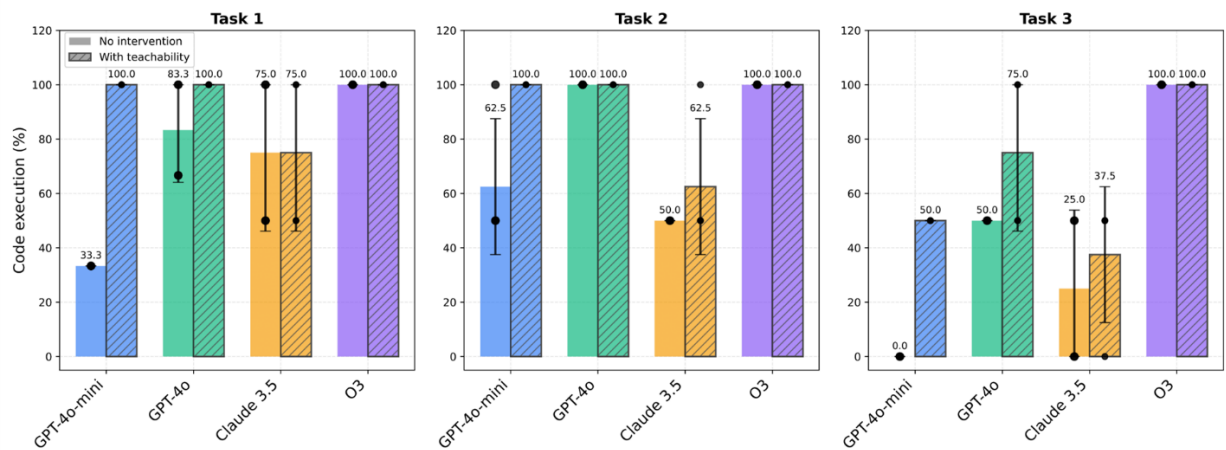

**Fig 4. Comparative evaluation of LLMs on operational tasks in X-ray beamline experiments**. The dashed bars represent model performance after incorporating and storing human feedback (teachability) whereas the solid bars represent their performance without any human intervention. The percentage on top of each bar shows the % performance in the accuracy of code execution. The dots represent the % performance of each experimental trial (4 trials in total) and the standard deviation is calculated from these four trials. The error bars are clipped to stay within 0-100% range. While human feedback improves performance in text-based tasks such as function calling or parameter specification, it has limited impact on enhancing the model's inherent visual reasoning capabilities.

## 2.4. Robotic polymer film fabrication

### 2.4.1 Robotic platform for thin film fabrication

To evaluate the generality of our agentic system, we deployed it in a distinct and more challenging context: the autonomous execution of complex, multi-step experimental workflows on a robotic platform for materials design. The platform consists of an N9 robotic arm (North Robotics) equipped with both finger and vacuum grippers, enabling the precise transfer of materials between modular workstations. Key stations include a vial rack for liquid precursors, a clamp holder for vial capping and uncapping, a substrate holder, a pipette tips rack, and a programmable blade-coating unit with tunable vacuum, temperature, and blade speed (Fig. 5a).

This system has previously been used in a range of materials design tasks ranging from metal films to conductive polymer films and devices.[28–30]

**2.4.2 Agentic pipeline evaluation at thin film fabrication platform**

For the N9 robotic station (Fig. 5), the main focus is on the coordination of the agents to plan and perform long sequential tasks where the correct order of the actions is critical for the success of the experiment. Routine operation of the robotic platform involves manually loading vials and substrates into their respective holders, after which pre-defined functions are typically called to carry out experimental procedures. To test our system in a more unconstrained and autonomous setting, we provided the agents only with basic low-level operation commands and the layout of the vials and substrates as a context (Fig 5b, Supplementary Fig. 6 and 7), removing access to high-level pre-defined routines. The agents were tasked with composing and sequencing these low-level commands to autonomously execute full experimental protocols. We designed three tasks of increasing complexity to benchmark system performance (Fig. 5c). A user-friendly interface with live-stream video capturing the N9 robot was also designed to enable the user-agent interaction (Supplementary Fig. 13).

The first task evaluated a simple action of transferring a polymer vial to the clamp holder using the finger gripper (Supplementary Fig. 12). The second task evaluated a medium complexity task of moving a substrate to the coating station via using the vacuum gripper and securely holding it there by activating the coating station vacuum whilst deactivating the vacuum gripper. The third and most complicated task required the full end-to-end fabrication of polymer thin film after retrieving the optimal processing conditions from a scientific paper which was provided in PDF format.[30] The third task integrates all previous steps into a coherent workflow (Fig 6b). The procedure included first using the literature scraper agent to read the PDF and identify the optimal parameters to create a polymer thin film of PEDOT:PSS without defects. The agent identified a coating temperature of $90°C$ and a coating speed of 1 mm/s as the optimal parameters to be further used for fabrication. The following steps included importing libraries, initializing hardware, identifying and selecting the correct polymer, retrieving and positioning a substrate via the Bernoulli vacuum gripper, and preparing the pipetting operation by uncapping the vial and aspirating the solution. The polymer solution was then drop-cast onto the substrate int the blade-coating station, where the temperature and coating speed parameters were set.

The agentic pipeline autonomously planned and executed this multi-step workflow, dynamically delegating subtasks among agents, generating and validating executable robotic code, and requesting human approval prior to execution to ensure operational safety. The integration of reasoning, planning, memory, and tool use, alongside literature-informed parameter extraction, enabled the system to complete Task 3 without hardcoded experimental scripts, demonstrating generalization beyond narrowly defined routines (Fig. 6a).

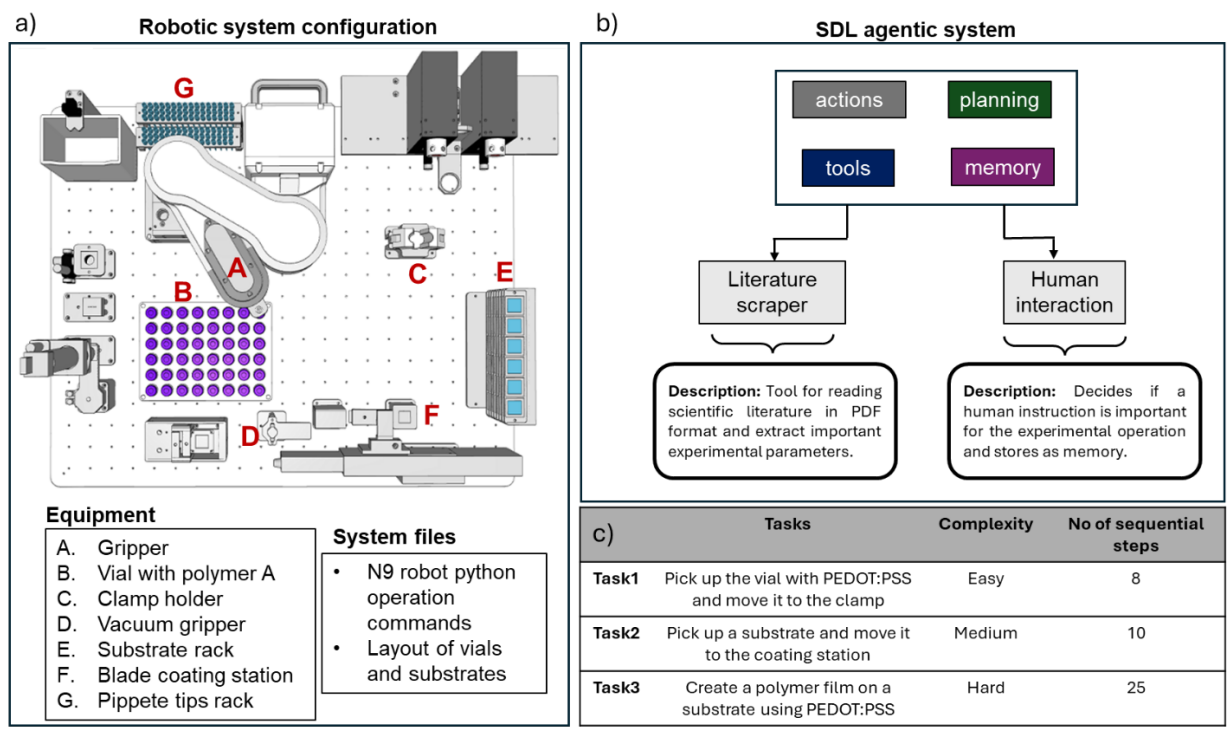

**Fig 5.** Autonomous robotic system and agentic pipeline for experimental design and execution. a) Schematic layout of the N9 robotic platform used for autonomous experimentation. The system includes a multi-functional robotic arm equipped with finger and vacuum grippers (A, D) for sample manipulation. Key modules include the vial rack containing liquid polymer samples (B), vial clamp holder for capping/uncapping (C), substrate rack (E), pipette tip rack (G), and a programmable blade-coating station (F) with tunable temperature, vacuum, and blade speed. Contextual system files, including Python command sets for robot control and the physical layout of vials and substrates, are provided to the agentic pipeline. b) Overview of the agentic pipeline, highlighting integration with external tools such as a literature scraper for extracting experimental parameters from scientific papers, and a human-interaction module for incorporating supervisory feedback and storing relevant memory. The agent framework includes capabilities for action execution, planning, tool use, and memory management. c) Summary of the evaluation tasks used to benchmark the pipeline. Tasks vary in complexity and number of required sequential steps.

Our evaluation revealed several key trends across the three experimental tasks (Fig. 6c). Without any human feedback, advanced language models such as GPT-4o and Claude-3.5 were capable of completing the relatively simple operations with few sequential steps (Task 1). However, as the complexity and number of actions increased in Tasks 2 and 3, all models showed a significant drop in performance in the zero-shot setting. Once human feedback was introduced through teachability, where agents received corrective demonstrations stored as input-output pairs in the memory database (Supporting Information Section 3), performance improved across all models. This suggests that memory augmented agent systems can generalize previously learned robotic workflows and reuse operational sequences effectively (Supplementary Section 4.2, Supplementary Tables 7-11).

Among the evaluated models, GPT-4o demonstrated high reliability and consistency in code execution across all tasks with teachability enabled. Claude 3.5, while performing well in simpler tasks, struggled with longer-horizon workflows and often failed to maintain correct sequencing without guidance. An interesting pattern was observed in the o3 reasoning model which frequently hallucinated commands and over-complicated even simple routines in Task 1, leading to execution failures. However, it showed a strong capacity for abstract reasoning in Task 3 when provided with human feedback, outperforming other models once the complexity increased. In addition, o3 often required excessive iterations between the code-writing and reviewing agents, introducing inefficiency in the execution pipeline. Overall, the introduction of memory through teachable, reusable feedback loops significantly enhanced agent performance by reducing redundant failures and reinforcing correct robotic control sequences. These results highlight the importance of integrating memory and correction-based learning in multi-agent systems for robust autonomous experimentation.

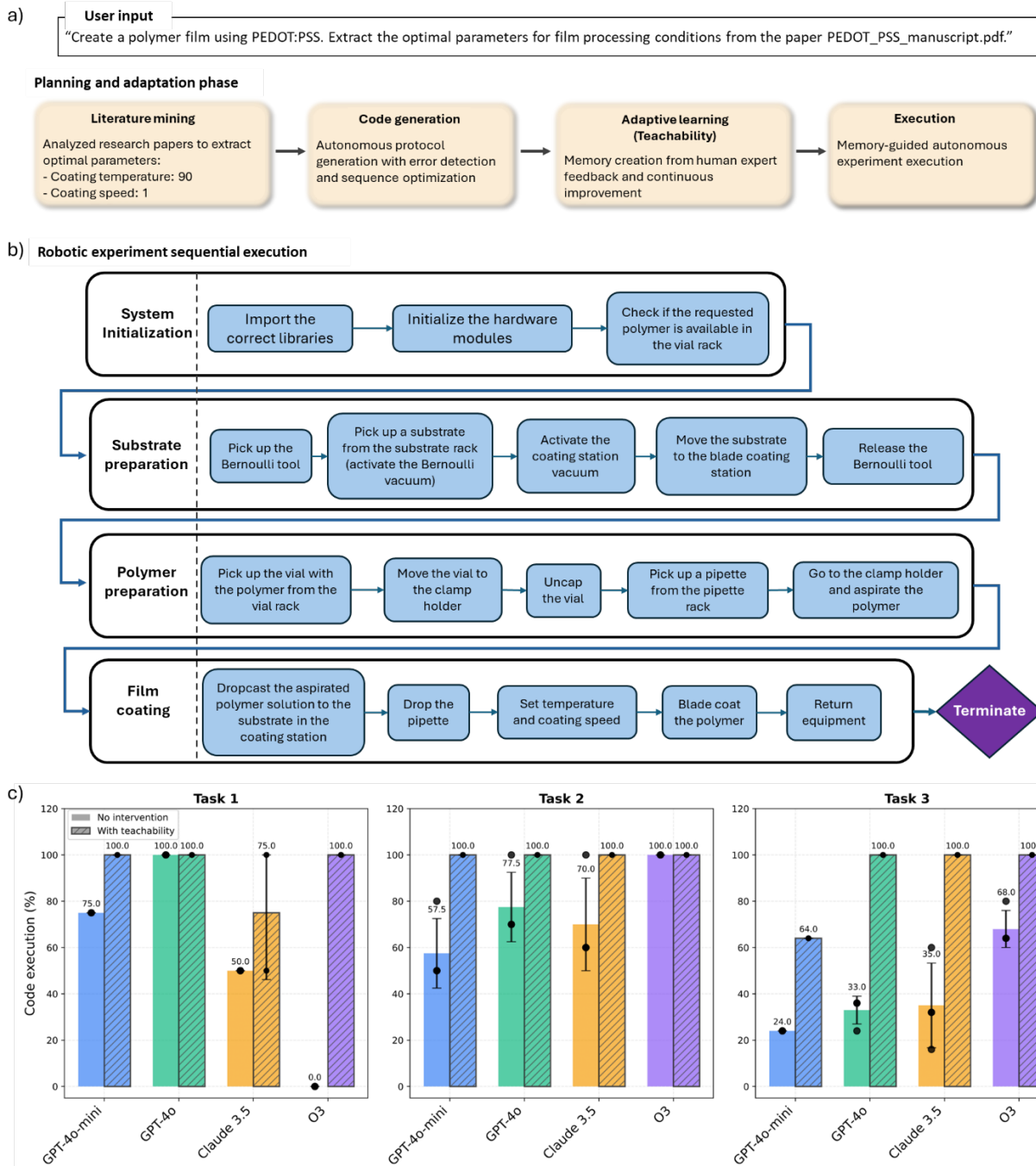

**Fig 6. Schematic of the agentic pipeline operating a highly complex set of robotic actions given a user prompt**. a) During the planning phase, the system had to first read a scientific paper and identify the conditions that led to a good quality film. The initial code was corrected with human feedback so that the agents will understand the correct sequence of steps to create the polymer film. b) Outline of the steps that the agentic system designed which include using the vacuum gripper to pick up an available substrate from the substrate rack, placing the substrate to the blade coating station and release the vacuum gripper, picking up the polymer from the vials rack and move it to the clamp, uncapping the clamp and aspirate the polymer with the pipette, drop

casting the polymer to the substrate and blade coat it. c) Comparative evaluation of LLMs on operational tasks for the N9 tasks. The error bars are clipped to stay within 0-100% range.

## 3. Summary and Outlook

In this work, we investigated the potential of AI agents to support complex, real-world scientific workflows. We developed and evaluated the capabilities and reliability of a multi-agent framework that leverages foundation models to autonomously operate robotized instrumentation in scientific user facilities such as beamlines and self-driving laboratories. In addition, we further investigated the impact of incorporating user guidelines into the system to enhance its ability to handle complex design and operational tasks.

Our results indicate that the agentic system is capable of learning from user interactions with the platform, refining its understanding of processes over time, and generating reliable code for robotic operations. This adaptability highlights the potential of trainable agents to improve the autonomy of self-driving laboratories, where human-in-the-loop interactions play a critical role in guiding and refining AI-driven workflows. While our teachable agent framework reduces the need for continual human intervention by storing and reusing past instructions, complete automation in real-world laboratory and user facility settings requires further safeguards. In particular, instrument drift and other hardware variances require regular calibration to maintain reliability over long experimental cycles. Likewise, reconfigurable workflows may introduce new instruments or procedures that must first be taught before running autonomously. Despite these challenges, once the core pipeline has been aligned with the lab environment and a sufficient body of memorized instructions has been built up, the agent can execute the tasks with minimal oversight. In our experiments, agents such as GPT-4o and Claude3.5-Sonnet were relatively easy to teach, whereas GTP-4o-mini required more iterative feedback to achieve an equivalent level of understanding. Whereas teachability can have a significant impact in explaining textual tasks such as function calling and leaning the correct sequential steps, it was observed that it cannot impact on enhancing the model's inherent visual reasoning capabilities.

Automation of protocols in an autonomous laboratory requires advanced models that can manage the sequential nature of tasks, perform quality checks, and provide real-time feedback to users. As multi-agent paradigms continue to evolve, they are expected to drive significant advancements in autonomous research, adaptive experimentation, and AI-guided discovery. Our results indicate that agentic systems can serve as a valuable resource for building self-driving facilities with enhanced autonomy, capable of learning from human expertise and improving over time.

## 4. Methods

**4.1 Agentic framework evaluation**

The evaluation metrics are categorized as follows: i) Code quality check: verifying that the agent performs code quality analysis for potential errors, stylistic inconsistencies, and best-practice violation. ii) Code correctness: evaluating the functionality of the generated code and checking whether the sequential steps are performed in the correct order. iii) Code execution: evaluating the functional correctness of the code, basically measuring in how many steps it will stop working in the real robotic environment, iv) Code repeatability: evaluating whether the same code is generating after running the same prompt three times. v) Code reproducibility: evaluating whether the generated code is the same after modifying the task prompts whilst keeping the main task the same. The results for each of the benchmarks are described in detail in Supporting Information, Section 4.

**4.2 Memory creation and teachability**

Autogen has a teachability feature that can add memory capabilities to any existing agent. By analyzing the incoming messages from the interaction with the human user, it detects whether the message contains general information that should be remembered or a task/problem with accompanying advice. If any of these are found, it extracts and generalizes the content, storing it as input-output pairs in the ChromaDB vector database. The input-output pairs are stored as text embeddings enabling semantic similarity searches. When a new task is provided by the user, the teachable agents perform a similarity search to find the most relevant past teachings. In that way, the agent can automatically apply relevant past learnings to new situations. The default teachability agent is storing all the created memories, even if they are very similar. To avoid storing very similar memories, we added a similarity search before storage. The new memories are compared to the existing ones using ChromaDB's similarity search, and a new memory is only stored if it is above a certain distance threshold. Using the memory component can also enable for the training and memory creation with more expensive Language models and operate with open-source and smaller ones.

## Data availability

All source data are provided within this paper. All relevant data are provided at the GitHub repository: [https://github.com/AdvancedPhotonSource/CALMS/tree/sdl_agents](https://github.com/AdvancedPhotonSource/CALMS/tree/sdl_agents).

## Code availability

All the relevant code associated with this work including are publicly available via GitHub at [https://github.com/AdvancedPhotonSource/CALMS/tree/sdl_agents](https://github.com/AdvancedPhotonSource/CALMS/tree/sdl_agents).

# Acknowledgments

Work was performed at the Center for Nanoscale Materials and Advanced Photon Source, both U.S. Department of Energy Office of Science User Facilities, supported by the U.S. DOE, Office of Basic Energy Sciences, under Contract No. DE-AC02-06CH11357.

## Author Contributions

AV and MHP worked on software development. TZ provided the workflows and nanoscale images for the X-ray Nanoprobe experiment. HC and AV provided the workflow and code for operating the N9 robotic station. AV and MJC conceived the project. All authors reviewed and commented on the manuscript.

## Competing interests

The authors declare no competing interests.

# References

1. Fernando, C. *et al.* Robotic integration for end-stations at scientific user facilities. *Digital Discovery* **4**, 1083–1091 (2025).

2. Kaiser, J., Lauscher, A. & Eichler, A. Large language models for human-machine collaborative particle accelerator tuning through natural language. *Sci Adv* **11**, eadr4173 (2025).
3. Song, T. *et al.* A Multiagent-Driven Robotic AI Chemist Enabling Autonomous Chemical Research On Demand. *J Am Chem Soc* **147**, 12534–12545 (2025).
4. OpenAI. OpenAI. Preprint at https://openai.com (2024).
5. Luo, X., Rechardt, A., Sun, G. & et al. Large language models surpass human experts in predicting neuroscience results. *Nat Hum Behav* (2024) doi:10.1038/s41562-024-02046-9.
6. Editorial. Prepare for truly useful large language models. *Nat Biomed Eng* **7**, 85–86 (2023).
7. Prince, M. H., Chan, H., Vriza, A. & others. Opportunities for retrieval and tool augmented large language models in scientific facilities. *NPJ Comput Mater* **10**, 251 (2024).
8. Van Herck, J. *et al.* Assessment of fine-tuned large language models for real-world chemistry and material science applications. *Chem. Sci.* **16**, 670–684 (2025).
9. Mathur, S., der Vleuten, N. van, Yager, K. G. & Tsai, E. H. R. VISION: a modular AI assistant for natural human-instrument interaction at scientific user facilities. *Mach Learn Sci Technol* **6**, (2025).
10. Hung, L. *et al.* Autonomous laboratories for accelerated materials discovery: a community survey and practical insights. *Digital Discovery* vol. 3 1273–1279 Preprint at https://doi.org/10.1039/d4dd00059e (2024).
11. Canty, R. B., Koscher, B. A., McDonald, M. A. & Jensen, K. F. Integrating autonomy into automated research platforms. *Digital Discovery* vol. 2 1259–1268 Preprint at https://doi.org/10.1039/d3dd00135k (2023).
12. Ruan, Y., Lu, C., Xu, N. & others. An automatic end-to-end chemical synthesis development platform powered by large language models. *Nat Commun* **15**, 10160 (2024).
13. Birhane, A., Kasirzadeh, A., Leslie, D. & Wachter, S. Science in the age of large language models. *Nature Reviews Physics* **5**, 277–280 (2023).
14. White, A. D. *et al.* Assessment of chemistry knowledge in large language models that generate code. *Digital Discovery* **2**, 368–376 (2023).
15. Bran, M., Cox, A., Schilter, S. & others. Augmenting large language models with chemistry tools. *Nat Mach Intell* **6**, 525–535 (2024).
16. Ansari, M., Watchorn, J., Brown, C. E. & Brown, J. S. dZiner: Rational Inverse Design of Materials with AI Agents. Preprint at https://doi.org/10.48550/arXiv.2410.03963 (2024).
17. Ma, K. AI agents in chemical research: GVIM – an intelligent research assistant system. *Digital Discovery* (2025) doi:10.1039/D4DD00398E.
18. Mathur, S., van der Vleuten, N., Yager, K. & Tsai, E. VISION: A Modular AI Assistant for Natural Human-Instrument Interaction at Scientific User Facilities. *arXiv preprint arXiv:2412.18161* (2024).
19. Darvish, K. *et al.* ORGANA: A robotic assistant for automated chemistry experimentation and characterization. *Matter* **8**, 101897 (2025).
20. langchain-ai. LangGraph.
21. Wu, Q. *et al.* AutoGen: Enabling Next-Gen LLM Applications via Multi-Agent Conversation. in *COLM 2024* (2024).
22. OpenAI. Swarm Agents.

23. Yin, X., Shi, C., Han, Y. & Jiang, Y. PEAR: A Robust and Flexible Automation Framework for Ptychography Enabled by Multiple Large Language Model Agents. *ArXiv* **abs/2410.09034**, (2024).
24. Muhoberac, M. *et al. State and Memory Is All You Need for Robust and Reliable AI Agents*.
25. Gao, S. *et al.* Empowering biomedical discovery with AI agents. *Cell* **187**, 6125–6151 (2024).
26. Ghafarollahi, A. & Buehler, M. J. AtomAgents: Alloy design and discovery through physics-aware multi-modal multi-agent artificial intelligence. (2024).
27. Winarski, R. P. *et al.* A hard X-ray nanoprobe beamline for nanoscale microscopy. *J Synchrotron Radiat* **19**, 1056–1060 (2012).
28. MacLeod, B. P. *et al.* Self-driving laboratory for accelerated discovery of thin-film materials. *Sci Adv* **6**, (2020).
29. MacLeod, B. P. *et al.* A self-driving laboratory advances the Pareto front for material properties. *Nature Communications 2022 13:1* **13**, 1–10 (2022).
30. Wang, C. *et al.* Autonomous platform for solution processing of electronic polymers. *Nat Commun* **16**, 1498 (2025).